\documentclass[12pt]{article}

\usepackage{scicite}

\usepackage{times}

\usepackage[inkscapeformat=png]{svg}
\usepackage{caption}
\usepackage{svg}
\usepackage{graphicx}
\usepackage{multirow}%
\usepackage{amsmath,amssymb,amsfonts}%
\usepackage{amsthm}%
\usepackage{mathrsfs}%
\usepackage[title]{appendix}%
\usepackage{xcolor}%
\usepackage{textcomp}%
\usepackage{manyfoot}%
\usepackage{booktabs}%
\usepackage{algorithm}%
\usepackage{algorithmicx}%
\usepackage{algpseudocode}%
\usepackage{listings}%
\usepackage{gensymb}

\newenvironment{sciabstract}{%
\begin{quote} \bf}
{\end{quote}}

\title{Tracking Surface Charge Dynamics on Single Nanoparticles}

\author
{\\Ritika Dagar,$^{1,2\ast\dag}$\,
    Wenbin Zhang,$^{1,2,3\ast\dag}$\,
    Philipp Rosenberger,$^{1,2}$
    \\Thomas M. Linker,$^{4}$\,
    Ana Sousa-Castillo,$^{5}$\,
    Marcel Neuhaus,$^{1,2}$\,
    Sambit Mitra,$^{1,2}$\,
    \\Shubhadeep Biswas,$^{1,2,4}$
    Alexandra Feinberg,$^{4}$\,
    Adam M. Summers,$^4$\,
    \\Aiichiro Nakano,$^{6}$
    Priya Vashishta$^{6}$\,
    Fuyuki Shimojo,$^7$,\ 
    Jian Wu,$^3$
    \\Cesar Costa Vera,$^{1,8}$\,
    Stefan A. Maier$^{9,10,}$\,
    Emiliano Cort\'es$^5$ 
    \\Boris Bergues$^{1,2,}$\,
    Matthias F. Kling$^{1,2,4,11\ast}$\\
\\\normalsize{$^{1}$Department of Physics, Ludwig-Maximilians-Universit{\"a}t Munich, D-85748 Garching, Germany}\\
    \normalsize{$^{2}$ Max Planck Institute of Quantum Optics,
    D-85748 Garching, Germany}\\
    \normalsize{$^{3}$State Key Laboratory of Precision Spectroscopy,
    East China Normal}\\ \normalsize{University Shanghai, 200241, China}\\
    \normalsize{$^{4}$Stanford PULSE Institute, SLAC National Accelerator Laboratory, Menlo Park, CA, 94025, USA}\\
    \normalsize{$^{5}$Nanoinstitute Munich, Department of Physics,}\\ \normalsize{
    Ludwig-Maximilians-Universit{\"a}t Munich, D-80539 Munich, Germany}\\
    \normalsize{$^{6}$Collobratory for Advanced Computing and Simulations, University of
    }\\ \normalsize{Southern California, Los Angeles, CA 90089, USA}\\
    \normalsize{$^{7}$Department of Physics, Kumamoto University, Kumamoto 860-0862,
    Japan}\\
    \normalsize{$^{8}$Department of Physics, Escuela Politecnica Nacional, Quito 170525,
    Ecuador}\\
    \normalsize{$^{9}$Department of Physics, Imperial College London, London SW7 2AZ, UK}\\
    \normalsize{$^{10}$School of Physics and Astronomy, Monash University,}\\ \normalsize{ Clayton Victoria 3800, Australia}\\
    \normalsize{$^{11}$Applied Physics Department, Stanford University, Stanford, CA,
    94305, USA}\\
    \normalsize{$^\ast$To whom correspondence should be addressed;}\\ 
    \normalsize{E-mail: ritika.dagar@physik.uni-muenchen.de, wbzhang@lps.ecnu.edu.cn,}\\ 
    \normalsize{ mfkling@slac.stanford.edu}\\
    \normalsize{$^\dag$These authors contributed equally to this work.}}

\date{}

\begin{document} 


\baselineskip18pt


\maketitle


\begin{sciabstract}
  Surface charges play a fundamental role in physics and chemistry, particularly in shaping catalytic properties of nanomaterials. Tracking nanoscale surface charge dynamics remains challenging due to the involved length and time scales. Here, we demonstrate real-time access to the nanoscale charge dynamics on dielectric nanoparticles employing reaction nanoscopy. We present four-dimensional visualization of the non-linear charge dynamics on strong-field irradiated single SiO$_2$ nanoparticles with femtosecond-nanometer resolution and reveal how surface charges affect surface molecular bonding with quantum dynamical simulations. We performed semi-classical simulations to uncover the roles of diffusion and charge loss in surface charge redistribution process. Understanding nanoscale surface charge dynamics and its influence on chemical bonding on a single nanoparticle level unlocks an increased ability to address global needs in renewable energy and advanced healthcare.
\end{sciabstract}



Altering charges in catalysts, surfaces, and even at the atomic level has been pivotal in numerous applications across diverse fields. Understanding the dynamics of surface charges holds significant promise for advancing atomic-scale technologies \cite{Repp2004} and enhancing our grasp of electrochemical reactions \cite{Carina2023}, catalyst selectivity \cite{Gallagher2023}, and surface adatom behavior \cite{Harutyuntan2020}. Surface charge manipulation in catalytic systems is an emerging field offering enhanced performance. Engineering surface charge states can endow unique functions upon catalytic nanosystems in terms of reactant adsorption and activation \cite{Bai2017}. One strategy to engineer surface charge states involves manipulating the surface electron density, such as through strong-field ionization processes that induce surface charges \cite{Seiffert2018}. 

Manipulating light at the nanoscale to generate charge carriers can enhance or initiate catalytic reactions through mechanisms like surface potential alteration, charge transfer, localized heating, and reactant binding \cite{Linic2015, Cortes2018}. Unlike plasmonic metal nanoparticles (NPs), which incur significant energy losses and heat generation when interacting with light, dielectric NPs provide an energy-efficient alternative \cite{JXu2021}. Recent research using strong-field ionized dielectric silica NPs has demonstrated their catalytic role in a photochemical reaction forming H$_3^+$ from water \cite{Alghabra2021}. This process relies on inducing charges on the NP's surface through the process of strong-field ionization. The interaction of strong-field lasers with the NPs leads to highly positively charged NPs, releasing energetic electrons \cite{Seiffert2017}. Electrons that are unable to escape the strong positive potential at the surface get trapped in the nanoscale environment and with local charge interactions, contribute to new reaction pathways \cite{Alghabra2021}. In spite of these advancements, establishing a precise relationship between dynamics of surface charges and the catalytic efficacy of NPs remains an open challenge, underscoring the need for further investigation.

Despite the significance of charge carriers in modifying photocatalytic reaction pathways, their intriguing ultrafast dynamics have yet to be fully elucidated. Ultrafast processes related to charge carriers include electron scattering \cite{Zherebtsov2011}, nanoplasma expansion \cite{Gorkhover2016,Peltz2022}, charge regulation \cite{Curk2021,Bakhshandeh2019}, and charge diffusion \cite{Xu2018am, Kim2018,Dizdar2018, He2017,Widenhorn2010, Bai2021,Zhou2013,Kozák2015,  Scajev2011pssa}. Unwrapping these processes on their natural timescales is critical for the development of next-generation solar harvesting devices \cite{Gupta2017}. Prior studies characterized surface charge generation and redistribution processes using methods such as atomic force microscopy \cite{Bai2021, Zhou2013} and optical light diffraction based on laser-induced transient grating techniques \cite{Kozák2015, Scajev2011pssa}. However, these approaches either necessitated the sample to be deposited on a substrate (affecting the surface charge dynamics and offering limited time resolution \cite{Bai2021, Zhou2013}) or are confined to macroscopic specimens (lacking spatial resolution \cite{Kozák2015, Scajev2011pssa}). 

Beyond a solid-state physics approach, a comprehensive understanding of how photo-induced surface charges affect the bonding motifs of surface functional groups is lacking in the literature for isolated nanosystems. Here, we present an approach to visualize surface charge dynamics on the surface of individual, isolated silicon dioxide (SiO$_2$) NPs subjected to infrared (IR) irradiation, in real time. Employing time-resolved reaction nanoscopy \cite{Rupp2019nc}, we uncover the ultrafast generation and redistribution of surface charges and demonstrate their influence on the terminal functional groups at the vacuum-NP interface.

\begin{figure}[t!]
\centering
\includegraphics[width=0.95\textwidth]{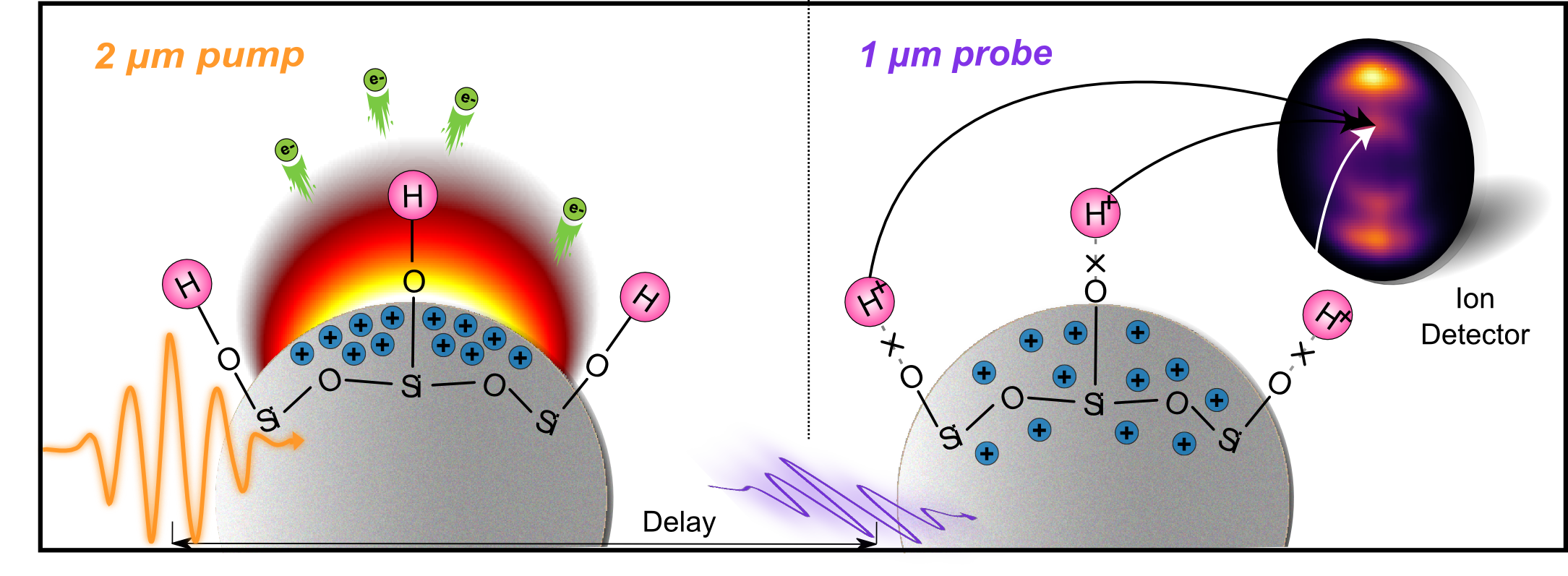}
\caption{\textbf{The pump-probe scheme for probing surface charge dynamics on SiO$_2$ NPs.}  Surface charges are generated on a silica NP by an infrared (IR) pump pulse,  which ionizes the NP through intensified near-fields, releasing electrons and resulting in the creation of positive surface charges (illustrated by blue plus signs), as depicted in the left panel. The presence of the surface charge weakens the terminal O-H bonds at the interface by delocalizing the electron density under the influence of the strong positive potential on the surface. The probe pulse, polarized perpendicularly to the pump pulse, dissociates the weakened bonds, resulting in the emission of protons (H$^+$ ions) in the polarization direction of the pump pulse. The protons' momentum distribution is utilized to spatially map the proton emission from the surface and, consequently, the distribution of surface charges as a function of the pump-probe delay.}
\end{figure} 

\vspace{18pt}
\textbf{Experimental Results.} We employed the reaction nanoscopy technique in conjunction with a two-color pump-probe scheme (see Fig. 1). Utilizing three-dimensional ion momentum spectroscopy, the reaction nanoscope enables the visualization of near-field-induced reaction yield distributions on isolated NP surfaces \cite{Rosenberger2022EJPD, Zhang2022OPTICA}. The technique relies on a point-projection approach that maps ion momentum distributions to real-space emission from the nanosurface. In the experiments described here, as depicted in Fig. 1, an IR pump pulse was employed to ionize the NP, resulting in the release of electrons and the generation of positive surface charges on individual SiO$_2$ NPs. Following a delay, a probe pulse induces the dissociation of ions from the terminal groups, such as physisorbed and chemisorbed solvent (ethanol) molecules (Si-O-CH$_2$CH$_3$) as well as O-H groups from silanols on the particle surface \cite{Zhang2022OPTICA}. The emitted ion fragments act as sensitive markers of the local chemical reaction landscape, where the pump-pulse-induced surface charges play a dominant role in influencing bond dissociation. Here, we used protons to probe the surface charge dynamics owing to their ubiquity among all emitting ions.

\begin{figure}[t!]
\centering
\includegraphics[width=0.85\textwidth]{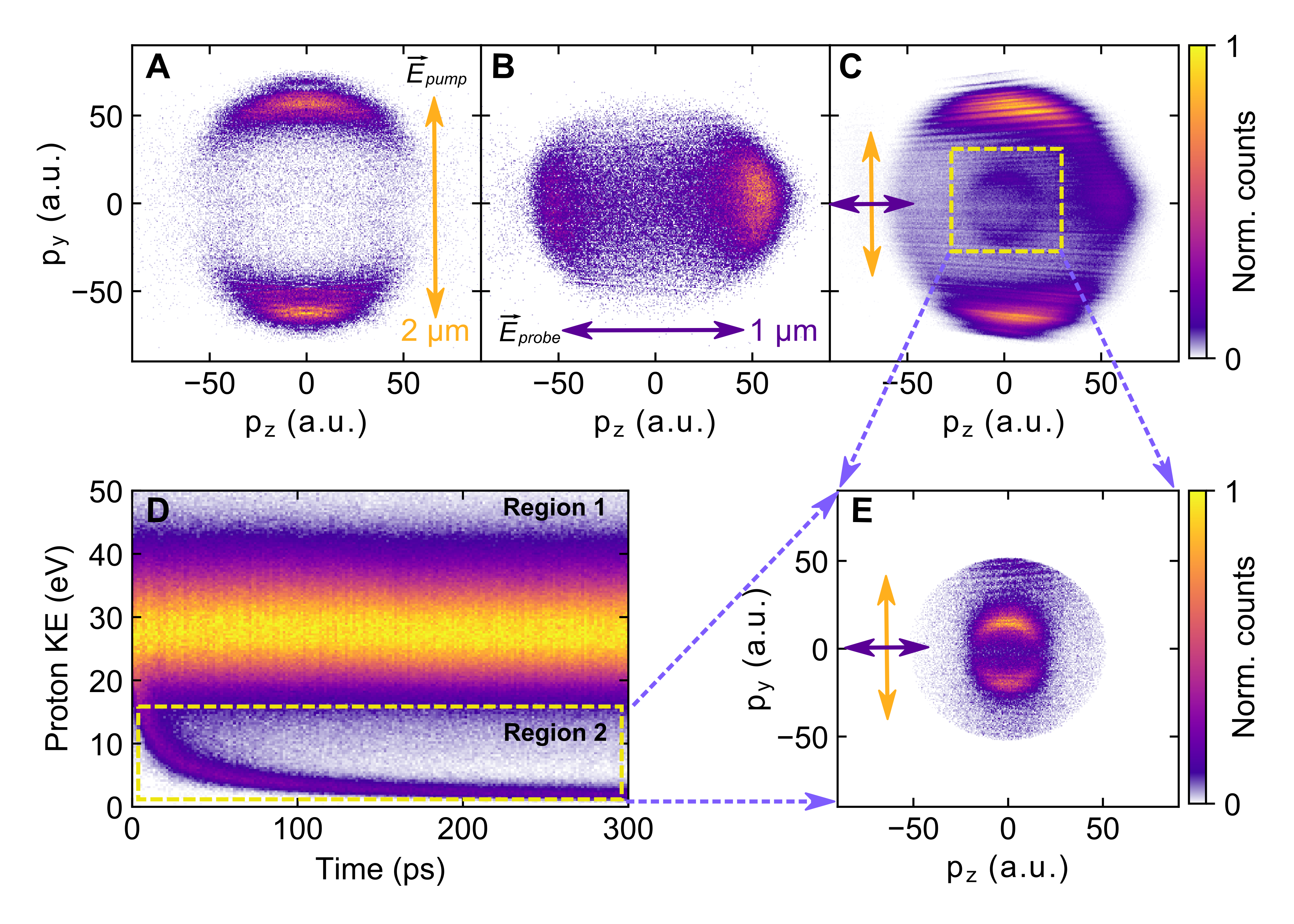} 
\caption{\textbf{Proton momentum and kinetic energy distributions.} Proton momentum distribution in the laser polarization plane measured by an individual (\textbf{A}) y-polarized 2 \textmu m pulses and (\textbf{B}) z-polarized 1 \textmu m pulses. ($\textbf{C}$) The time-integrated proton momentum distribution ($p_x$ = 0 plane) when both 2 \textmu m and 1 \textmu m pulses are introduced, where the 2 \textmu m pulses serve as the pump and 1 \textmu m pulses serve as the probe. An additional feature at lower momentum is identified within the yellow-dashed square, distinct from the individual contributions of 2 \textmu m and 1 \textmu m pulses (A, B). ($\textbf{D}$) Proton kinetic energy spectrum measured as a function of time delay between 2 \textmu m and 1 \textmu m pulses (2 \textmu m pulses preceding 1 \textmu m pulses). ($\textbf{E}$) The proton momentum distribution associated with the time-dependent KE decay signal in Region 2 of (D) coincides with the inner ring structure marked by the dashed square in panel (C).}
\end{figure}

The experimental framework involved generating an aerosolized beam of spherical dielectric SiO$_2$ NPs using an atomizer and aerodynamic lens system. The aerodynamic focusing system allows for the investigation of single NPs and ensures a fresh sample for each laser shot. The pump consisted of laser pulses centered around the wavelength of 2 \textmu m, with pulse duration of $\sim$ 35 fs at an intensity of $\sim$ 2 × 10$^{13}$ W/cm$^2$. The second harmonic of the pump pulse at 1 \textmu m, with a pulse duration of $\sim$ 40 fs and intensity $\sim$ 3.8 × 10$^{12}$ W/cm$^2$, polarized perpendicularly to the pump pulse, was used as the probe. The orthogonal selection of the polarization direction for pump-probe pulses was vital for distinguishing the unique interaction of individual pulses with the NP surface leading to distinct optical and chemical hotspots on the NP surface for each pulse. We employed moderate laser intensities to suppress plasma formation \cite{Hickstein2014acs} and particle expansion \cite{Gorkhover2016, Peltz2022} upon irradiation. The distinct abundance of released
electrons, detected with a channeltron detector was used to differentiate NP events from background gas. A time- and position-sensitive detector reconstructed the 3D-ion-momentum distribution. More experimental details can be found in the Supplemental Material.

Figure 2 represents the measured proton momentum and kinetic energy (KE) distributions for 300 nm SiO$_2$ nanospheres. Figures 2A and 2B show dipolar momentum distributions observed from the pump and probe pulses separately, i.e., 2 \textmu m pump and 1 \textmu m probe pulses along their polarization in the y- and z- directions (defined in the lab frame), respectively. Such features agree with previous experimental results \cite{Rupp2019nc, Rosenberger2020, Zhang2022OPTICA, Rosenberger2022EJPD} in which the ion momenta distribution closely follows the laser-polarization-sensitive local field enhancement around the NP. In our work, we identified an additional feature at lower momentum, contained by the dashed box (see Fig. 2C) that depicts the time-integrated proton momentum distribution in the shared polarization plane of both pulses, i.e., $p_x$ = 0 plane. Figure 2D displays the proton KE spectrum as a function of the pump-probe delay. The spectrum can be separated into two parts: proton energies above and below 20 eV. The high-energy band ($\sim$ 28 eV) is delay-independent, which mainly originates from individual pump or probe pulses. However, the energy region below 20 eV shows a clear time dependence where the proton KE decreases with increasing pump-probe delay. Starting from the same energy region as the single-pulse-induced high-energy band ($>$ 20 eV in Fig. 2D), the proton KE decreases as a function of time delay, to an asymptotic energy of $\sim$ 2.3 eV within approximately 300 ps. 

Focusing our analysis on the low-KE region of the proton momentum distribution, corresponding to the inner momentum ring in Fig. 2C, one observes that protons are emitted along the polarization direction of the pump pulse. This is observed even though the probe is perpendicularly polarized to the pump, and is primarily responsible for triggering proton emission. We observed consistent results for the negative delays in which the 1 \textmu m pulse serves as the pump (see Fig. S3). Our investigation further revealed that a single exponential curve proved inadequate in accurately representing the time-dependent decay of KE observed in the experiments. As a result, we employed a bi-exponential model to better capture and model the observed decay trace. Two distinct time constants were derived from the fitting, a faster process occurring around the timescale of $\sim$ 10 ps, while the slower process exhibits decay over a longer duration with a time constant of roughly 65 ps (refer to Fig. S4). This implies that the delay-dependent proton emission involves a two-step process. At first, positive surface charges are generated by the initial pump pulse whose distribution aligns with the laser's polarization-sensitive near-fields. The interaction between surface charges created by the pump pulse and the highly electronegative surface oxygen atom of the NPs prompts an induced polarization within the surface bonds of the NPs. This induced polarization leads to a  redistribution of the electron density within the surface bonds and leads to bond weakening, further corroborated by our quantum dynamical simulations as discussed later. Since the surface molecular bonding has been weakened by the adjacent local surface charges, in the second step, the later-arriving, perpendicularly polarized probe pulse can more readily yield surface molecular dissociation with the generation of protons closely correlating with the surface charge distribution. In the course of dissociation, the surface charges attract electron density away from the dissociating surface groups, leading to a reduction in the Coulombic potential across the NP surface. Therefore, the Coulomb potential experienced by the protons is reduced, leading to a slowdown of the KE decay after a few tens of picoseconds. A similar feature is observed with H$_2^+$ emission (see Fig. S6).

\begin{figure}
\centering
\includegraphics[width=1\textwidth]{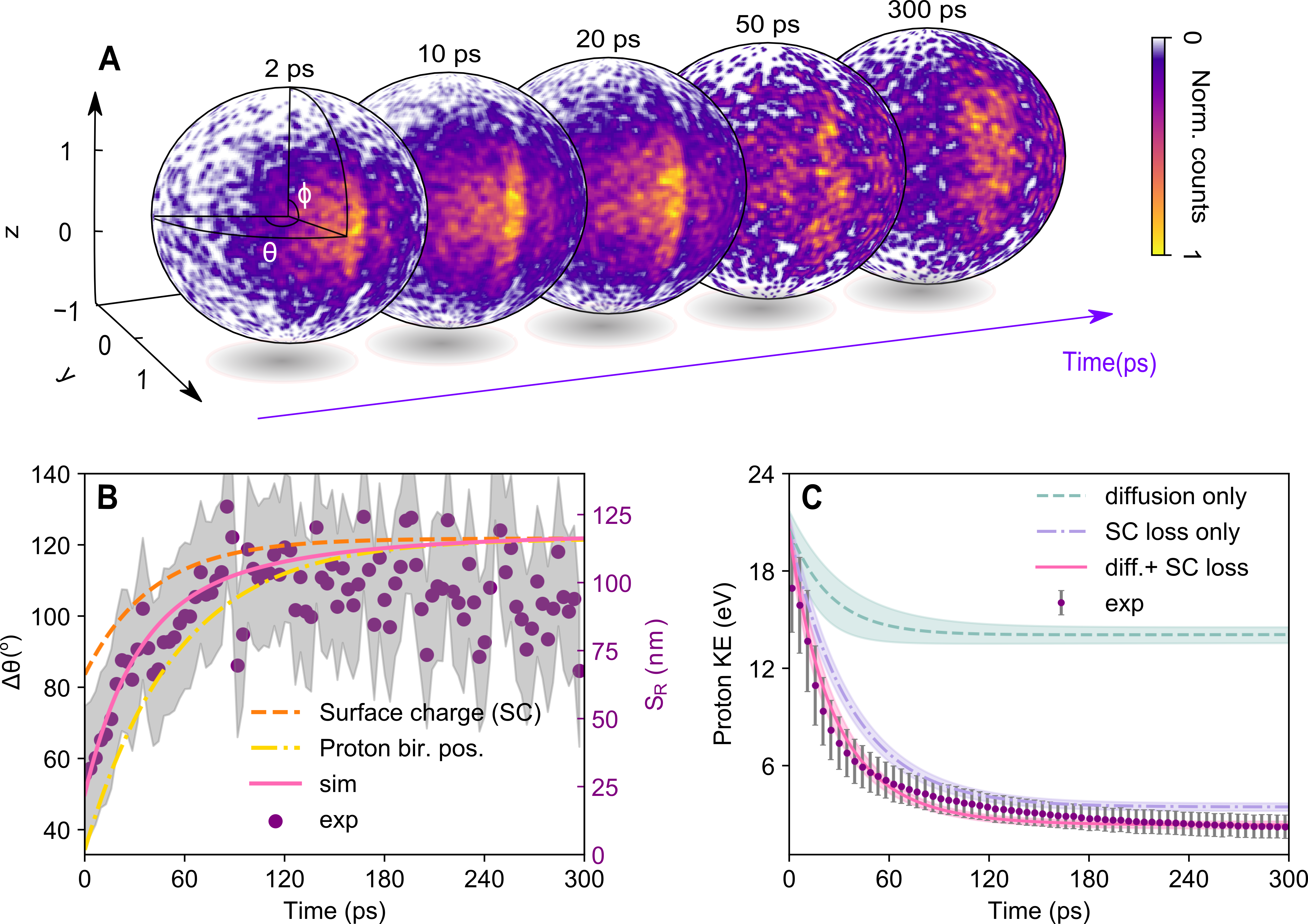}
\caption{\textbf{Angular proton momentum distribution} (\textbf{A}) Snapshots of the 3D angular proton momentum distribution at various time delays, reflecting changes in the spread of the surface charge distribution over time. The 3D momentum distributions of protons ($\phi$, $\theta$, $r$) are integrated radially and projected onto a unit sphere as a two-dimensional density map ($\phi$, $\theta$). The sphere coordinates are defined by elevation ($\phi$) and azimuthal ($\theta$) angles within specific intervals, where $\theta$ spans  $[-\pi, \pi]$ relative to the x-axis (propagation direction), and $\phi$ spans ($[-\pi/2, \pi/2]$ in relation to the y-z plane (pump polarization - TOF). The color scale represents the proton count within each angle. (\textbf{B}) The variation in the delay-dependent full-width half maximum values of the angular distribution $\theta$ in A, represented by $\Delta\theta$, as extracted from the Gaussian fits for the time-sliced theta distributions. The gray shaded area depicts the standard errors from the fittings. The angular broadening of the simulated angular distributions of surface charges (orange dashed curve), proton birth position (yellow dashed-dot curve), and proton momentum distributions (pink solid curve) are also shown. (\textbf{C}) Mean proton KE values derived from the simulations with different set parameters as a function of time, along with the experimental values. The dashed curve shows proton KE decay over time in the case of surface charge diffusion only and charge loss only denoted by the dash-dotted line, and the solid line combines both cases. Purple dots with gray error bars indicate experimentally measured mean proton KE values with associated standard deviations.}
\end{figure}

As shown in Fig. 3, the proton momentum distributions in the low-KE region also exhibit an angular broadening. Figure 3A provides a visual representation of the 3D ($\phi$, $\theta$, $r$) angular proton momentum distribution at different time intervals, capturing the evolving pattern of the time-dependent momentum distribution of the protons exhibiting KE-decay. The variation in the distribution of $\theta$, i.e., $\Delta\theta$, to time, is depicted in Fig. 3B. The results elucidate an exponential redistribution of the surface charges accompanied by a wider distribution of proton birth positions and consequently proton momentum distribution. The proton angular broadening occurs more rapidly during the initial short-time delays (less than 30 ps) and slows down at later time delays (beyond 50 ps), eventually reaching a plateau over 100 ps. The $\Delta\theta$ increases swiftly within the first 30 ps, increasing roughly from 50$\degree$ to around 80$\degree$, accompanied by a proportional reduction in proton KE ($\sim$24 eV to $\sim$7.7 eV). The observed proton KE decreasing and angular broadening over time can be attributed to two factors. First, as the initially localized surface charge distribution spreads out, the emitted protons are expected to possess lower final proton KE due to the diminishing Coulomb force associated with the reduced surface charge density. Second, since the directionality of proton emission is closely correlated with the surface charge distribution, a broader angular distribution of protons is anticipated when the charge distribution itself broadens.

By translating the proton emission angle width to the spatial position of surface charges using the formula, $S_{R}(\theta) = \pi[(\Delta\theta (t) - \Delta\theta (t=0))/(2\times360{\degree})]d$, where $d$ is the diameter of the NP, one can obtain the time-dependent relative change in the position of charges to their birth position on the nanoparticle surfaces. The right-purple-axis in Fig. 3B denotes the relative change in proton birth positions, $S_R$, spatially on the 300 nm SiO$_2$ nanosphere. The calculation is based on the near one-to-one mapping between the initial proton birth position and final proton momentum \cite{Zhang2022OPTICA}. We trace the relative spatial movement of surface charges through the dissociation of protons caused by the presence of these surface charges. Therefore, our analysis, on one hand, uncovers how the surface charge redistribution dynamics affect the proton birth position and their final momentum, and on the other hand, underscores the crucial impact of the surface charge distribution on proton emission via surface bond weakening. This methodology enables us to track the spatiotemporal evolution of surface charge density with nanoscopic spatial resolution and femtosecond-temporal resolution.

\vspace{18pt}
\textbf{Theoretical results.}We employed two different simulation frameworks to understand the observed charge dynamics and the molecular bond weakening on the surface of SiO$_2$ NPs (see more details in the Supplemental Material). We first performed the diffusion-model-based classical trajectory Monte Carlo (d- CTMC) simulations to understand the process of surface charge redistribution and then the non-adiabatic quantum molecular dynamics (NAQMD) simulations \cite{Fuyuki2013, Fuyuki2014} to demonstrate the role of local surface charges created by the pump pulse on bond weakening.

To evaluate the role of surface charge diffusion and loss of charge on the proton kinetic energy dynamics, we used the d-CTMC simulations. In this model, local positive surface charge distribution was generated on a defined-size spherical particle by quasi-static ionization, leaving point-like positively charged ions \cite{Rosenberger2020,Rosenberger2022EJPD,Zhang2022OPTICA}. These charges' initial distribution correlates with enhanced near-fields induced by the pump laser. A Gaussian random walk simulates the charge redistribution based on the diffusion process. The protons are propagated in the electrostatic field of redistributed charges in each time step. We then routinely calculated the final proton momentum spectra as a function of time. Our study explores the diffusion-based charge relaxation coupled with charge loss on proton dynamics. The simulation results qualitatively match experimental observations of time-resolved proton KE decay (Fig. 3C) and broadened proton angular momentum distributions (Fig. 3B). Individual impacts of the diffusion and charge loss processes are also examined (see Fig. 3C). The d-CTMC simulations yield significant key parameters, a diffusion coefficient of approximately  1.35 cm$^2$/s, and the charge decay constant, whose value is approximated to be around 40 ns, taking into account some amount of residual charge. The values of these constants that primarily rely on material characteristics are corroborated by assessing them with different SiO$_2$ NP sizes, as detailed in the Supplementary Material. 

In the d-CTMC simulations, it was postulated that the likelihood of proton dissociation from silanols at the NP surface is influenced by the presence of surface charges generated by the pump pulse, which, in turn, eases the weakening of these bonds and eventually leads to dissociation during their stretching by the probe pulse. To verify and quantify this, we used the NAQMD simulations on an amorphous silica surface terminated with Si-OH and Si-OH$_2$ groups in the presence of excited holes from the valence band maximum (VBM). Figure 4A shows the simulation slab replicated, the two-white lines representing the periodic boundary condition. We examined the bond dynamics for the O-H bond on the surface for which the hole wave function had the highest Mulliken population. Figure 4B shows the bond dynamics for the O-H bond with and without the presence of any holes. The excitation of a single hole from the VBM weakens the bond resulting in longer stretching length and lower stretching frequency. If the VBM becomes doubly unoccupied the O-H bond quickly breaks. The initial pump pulse is anticipated to generate many holes that can quickly dissociate O-H bonds and leave the NP. The remaining holes that have not led to dissociation still weaken the O-H bond, which can be later broken by the probe pulse during the stretching of the bonds. This underscores the crucial role of local surface charge in affecting surface adsorbate bonding by inducing bond weakening. 

\begin{figure}
\includegraphics[width=1\textwidth]{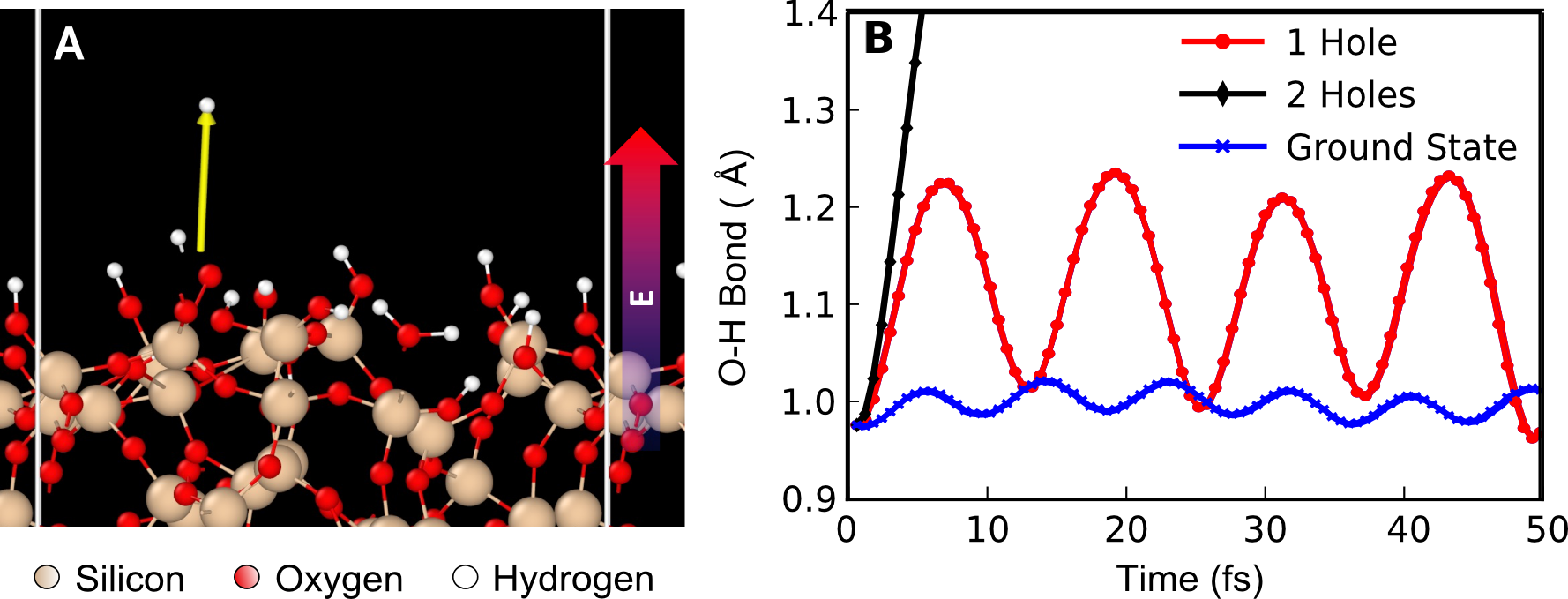} 
\caption{\textbf{Bond weakening due to the presence of surface charges}  (\textbf{A}) Simulated silica slab representing a $\sim$1\,nm$^2$ chunk of silica in the presence of a strong DC field generated by the surface charges. The white lines depict the implemented periodic boundary conditions. (\textbf{B}) O-H bond dynamics with and without the presence of holes.}
\end{figure}

 \subsubsection*{Discussion}
Our d-CTMC simulations indicate that the angular broadening observed in the proton momentum distribution (see Fig. 3B) primarily stems from diffusion, as it spreads initially localized surface charges over time due to its stochastic nature. However, simulations employing just diffusion towards a homogeneous distribution were unable to accurately model the entire proton KE decay profile even though they were able to reproduce the initial rapid decline. Implementing a charge loss mechanism driven by proton detachment, markedly enhanced the agreement between our simulations and the experimental data. Protons emitted through this mechanism will achieve a reduced final KE owing to the diminished Coulomb force resulting from the lower number and increased separation of the surface charges as a consequence of the considerable effect of electrostatic interactions from the charged NP surface. Examining the energy loss in protons allows us to probe the local and overall surface charge density on the NP surface. The initial ion momentum from molecular dissociation is considered negligible. Importantly, our study is not only able to comprehend the roles of both diffusion and charge loss in the spatiotemporal dynamics of surface charges, but is able to distinguish them based on the order of magnitude of their characteristic timescales. The ability to extract and differentiate the timescales of these processes on isolated NPs permits comparison to theoretical predictions, and ultimately drive a deeper understanding of NP-driven catalysis and related charge-mediated processes.

The calculated diffusion coefficient of around 1.35 cm$^2$/s in the simulations is indicative of the unique characteristics of the strong-field ionization of the SiO$_2$ NPs. While in comparison to a degenerate electron gas (i.e simple metal) such a diffusion constant would be small \cite{Palenskis2013},  but it is quite large for typically reported photocarrier transport coefficients \cite{Xiwen2019, Samuel2013, Kissel2022}, and in comparison to typical photocatalytic material TiO$_2$ our reported diffusion constant is three orders of magnitude larger than its bulk value \cite{Deskins2007}. The corresponding charge carrier mobility for our reported diffusion constant would be on the high end for those observed in heavily doped wide band gap semiconductors \cite{Kim2012}. As the trapping field arising from the positively charged surface and free electrons created by the strong field ionization of SiO$_2$ NPs leads to strong charge interactions and increased charge mobility at the surface \cite{Seiffert2017}, our reported diffusion constant, between that of a heavily doped semiconductor and a degenerate electron gas is consistent with this scenario. The charge decay constant, approximately 40 ns, also exhibits a degree of similarity with reported values observed in analogous systems. For instance, in rutile TiO$_2$, the lifetimes of holes were found to exceed several hundred picoseconds \cite{SHEN2006}. Development of next generation photocatalytic materials is extremely reliant on knowledge and ability to engineer charge carrier lifetime and diffusion properties, with typically long charge migration times sought to maximize reactivity \cite{Wang2020}. The implemented pump-probe reaction nanoscopy allows the evaluation of these properties at the nanoscale.


Our theoretical framework offers a highly intuitive interpretation of our experimental observations. We do acknowledge some limitations in our calculations, as there are additional processes, such as charge-charge interactions and surface scattering, which may influence the diffusion and charge loss process. By taking into account these complex factors, a more quantitative and extensive knowledge of the intricate functioning of nanomaterials can be achieved. Nevertheless, our simple approach combined with the experimental findings furnishes a thorough and clear understanding of the nanoscale redistribution dynamics of surface charges on individual NPs. This framework can provide a valuable means to comprehend diverse nanomaterials and more generally, charge-mediated processes on surfaces.

In conclusion, we demonstrated 4D visualization of surface charge dynamics on laser-irradiated dielectric silica NPs through pump-probe reaction nanoscopy. The semi-classical simulations manifested distinct timescales for the dispersion and loss of surface charges, yielding the experimentally observed bi-exponential decline in the kinetic energies of the dissociating protons. The NAQMD simulations revealed the strong influence of the surface charges on the bonding of surface functional groups, demonstrating the capability of charged dielectric NPs to trigger chemical reactions at the surface. Such dynamics are tracked via the probe and can be used to reconstruct charge diffusion and decay properties. In the context of silica NPs, these findings explain the peculiar catalytic reactivity previously observed in related studies \cite{Alghabra2021}. In a broader sense, the time-resolved experiments of charge dynamics on a single nanoparticle level can be used to investigate the role of particle material and morphology on charge carrier dynamics. Such understanding will be critical for optimizing charge-induced heterogeneous catalysis. 



\bibliography{scibib}

\begin{thebibliography}{10}

\bibitem{Repp2004}
J.~Repp, G.~Meyer, F.~E. Olsson, M.~Persson, {\it Science\/} {\bf 305}, 493 (2004).

\bibitem{Carina2023}
C.~Y.~J. Lim, {\it et~al.\/}, {\it Nature Communications\/} {\bf 14}, 335 (2023).

\bibitem{Gallagher2023}
J.~Gallagher, {\it Nature Energy\/} {\bf 8}, 112 (2023).

\bibitem{Harutyuntan2020}
H.~Harutyunyan, F.~Suchanek, R.~Lemasters, J.~J. Foley, {\it MRS Bulletin\/} {\bf 45}, 32 (2020).

\bibitem{Bai2017}
Y.~Bai, H.~Huang, C.~Wang, R.~Long, Y.~Xiong, {\it Materials Chemistry Frontiers\/} {\bf 1}, 1951–1964 (2017).

\bibitem{Seiffert2018}
L.~Seiffert, S.~Zherebtsov, M.~F. Kling, T.~Fennel, {\it Adv. Phys. X\/} {\bf 7}, 2010595 (2022).

\bibitem{Linic2015}
S.~Linic, U.~Aslam, C.~Boerigter, M.~Morabito, {\it Nature Materials\/} {\bf 14}, 567–576 (2015).

\bibitem{Cortes2018}
E.~Cort\'es', {\it Science\/} {\bf 362}, 28 (2018).

\bibitem{JXu2021}
J.~Xu, {\it et~al.\/}, {\it Advanced Optical Materials\/} {\bf 9}, 2100112 (2021).

\bibitem{Alghabra2021}
M.~S. Alghabra, {\it et~al.\/}, {\it Nat. Commun.\/} {\bf 12}, 3839 (2021).

\bibitem{Seiffert2017}
L.~Seiffert, {\it et~al.\/}, {\it Journal of Modern Optics\/} {\bf 64}, 1096 (2017).

\bibitem{Zherebtsov2011}
S.~Zherebtsov, {\it et~al.\/}, {\it Nat. Phys.\/} {\bf 7}, 656 (2011).

\bibitem{Gorkhover2016}
T.~Gorkhover, {\it et~al.\/}, {\it Nat. Photon.\/} {\bf 10}, 93 (2016).

\bibitem{Peltz2022}
C.~Peltz, {\it et~al.\/}, {\it New J. Phys.\/} {\bf 24}, 043024 (2022).

\bibitem{Curk2021}
T.~Curk, E.~Luijten, {\it Phys. Rev. Lett.\/} {\bf 126}, 138003 (2021).

\bibitem{Bakhshandeh2019}
A.~Bakhshandeh, D.~Frydel, A.~Diehl, Y.~Levin, {\it Phys. Rev. Lett.\/} {\bf 123}, 208004 (2019).

\bibitem{Xu2018am}
C.~Xu, {\it et~al.\/}, {\it Adv. Mater.\/} {\bf 30}, 1803968 (2018).

\bibitem{Kim2018}
S.~M. Kim, {\it Eur. Phys. J. Plus\/} {\bf 133}, 535 (2018).

\bibitem{Dizdar2018}
T.~O. Dizdar, G.~Kocausta, E.~Gülcan, Özcan Y.~Gülsoy, {\it Powder Technol.\/} {\bf 327}, 89 (2018).

\bibitem{He2017}
Y.~He, W.~Xie, Y.~Zhao, H.~Li, S.~Wang, {\it Int. J. Miner Process.\/} {\bf 166}, 7 (2017).

\bibitem{Widenhorn2010}
R.~Widenhorn, A.~W. Bargioni, M.~M. Blouke, A.~J. Bae, E.~Bodegom, {\it Optical Engineering\/} {\bf 49(4)}, 044401 (2010).

\bibitem{Bai2021}
X.~Bai, A.~Riet, S.~Xu, D.~J. Lacks, H.~Wang, {\it J. Phys. Chem. C\/} {\bf 125}, 11677 (2021).

\bibitem{Zhou2013}
Y.~Zhou, {\it et~al.\/}, {\it Nano Lett.\/} {\bf 13}, 2771 (2013).

\bibitem{Kozák2015}
M.~Kozák, F.~Trojánek, P.~Malý, {\it New J. Phys.\/} {\bf 17}, 053027 (2015).

\bibitem{Scajev2011pssa}
P.~Ščajev, V.~Gudelis, E.~Ivakin, K.~Jarašiunas, {\it Phys. Status Solidi A\/} {\bf 208}, 2067 (2011).

\bibitem{Gupta2017}
R.~Gupta, B.~Rai, {\it Sci. Rep.\/} {\bf 7}, 45292 (2017).

\bibitem{Rupp2019nc}
P.~Rupp, {\it et~al.\/}, {\it Nat. Commun.\/} {\bf 10}, 1 (2019).

\bibitem{Rosenberger2022EJPD}
P.~Rosenberger, {\it et~al.\/}, {\it Eur. Phys. J. D\/} {\bf 76}, 109 (2022).

\bibitem{Zhang2022OPTICA}
W.~Zhang, {\it et~al.\/}, {\it Optica\/} {\bf 9}, 551 (2022).

\bibitem{Hickstein2014acs}
D.~D. Hickstein, {\it et~al.\/}, {\it ACS Nano\/} {\bf 8}, 8810–8818 (2014).

\bibitem{Rosenberger2020}
P.~Rosenberger, {\it et~al.\/}, {\it ACS Photonics\/} {\bf 7}, 1885–1892 (2020).

\bibitem{Fuyuki2013}
F.~Shimojo, {\it et~al.\/}, {\it Comput. Phys. Commun.\/} {\bf 184}, 1 (2013).

\bibitem{Fuyuki2014}
F.~Shimojo, {\it et~al.\/}, {\it J. Chem. Phys.\/} {\bf 140}, 18A529 (2014).

\bibitem{Palenskis2013}
V.~Palenskis, {\it World Journal of Condensed Matter Physics\/} {\bf 3}, 73 (2013).

\bibitem{Xiwen2019}
X.~Gong, {\it et~al.\/}, {\it Nat Commun\/} {\bf 10}, 1591 (2019).

\bibitem{Samuel2013}
S.~D. Stranks, {\it et~al.\/}, {\it Science\/} {\bf 342}, 341 (2013).

\bibitem{Kissel2022}
M.~Kissel, {\it et~al.\/}, {\it Advanced Materials\/} {\bf 34}, 2203032 (2022).

\bibitem{Deskins2007}
N.~A. Deskins, M.~Dupuis, {\it Phys. Rev. B\/} {\bf 75}, 195212 (2007).

\bibitem{Kim2012}
H.~J. Kim, {\it et~al.\/}, {\it Applied Physics Express\/} {\bf 5}, 061102 (2012).

\bibitem{SHEN2006}
Q.~Shen, {\it et~al.\/}, {\it Chemical Physics Letters\/} {\bf 419}, 464 (2006).

\bibitem{Wang2020}
Q.~Wang, K.~Domen, {\it Chemical Reviews\/} {\bf 120}, 919 (2020).

\bibitem{Neuhaus2018}
M.~Neuhaus, {\it et~al.\/}, {\it Opt. Exp.\/} {\bf 26}, 16074 (2018).

\bibitem{Stober1968}
W.~Stöber, A.~Fink, E.~Bohn, {\it JOURNAL OF COLLOID AND INTERFACE SCIENCE\/} {\bf 26}, 62 (1968).

\bibitem{Ammosov1986}
M.~V. Ammosov, N.~B. Delone, V.~P. Krainov, {\it Sov. Phys. JETP\/} {\bf 64}, 1191–1194 (1986).

\bibitem{Perdew1996}
J.~P. Perdew, K.~Burke, , M.~Ernzerhof, {\it Phys Rev Lett\/} {\bf 77}, 3865 (1996).

\bibitem{Grimme2010}
S.~Grimme, J.~Antony, S.~Ehrlich, H.~Krieg, {\it J. Chem. Phys.\/} {\bf 132}, 154104 (2010).

\bibitem{Blochl1994}
P.~E. Blöchl, {\it Phys Rev B\/} {\bf 50}, 17953 (1994).

\bibitem{Tully1990}
J.~C. Tully, {\it J. Chem. Phys.\/} {\bf 93}, 1061 (1990).

\bibitem{Shimojo2019}
F.~Shimojo, {\it et~al.\/}, {\it SoftwareX\/} {\bf 10}, 100307 (2019).

\bibitem{Süßmann2015nc}
F.~Süßmann, {\it et~al.\/}, {\it Nat. Commun.\/} {\bf 92}, 5645–5649 (2015).

\bibitem{Ekanayake2018}
N.~Ekanayake, {\it et~al.\/}, {\it Nature Communications\/} {\bf 9}, 5186 (2018).

\end{thebibliography}

\bibliographystyle{Science}

\section*{Data availability}
All the data that support the findings of this study are available from the corresponding author upon reasonable request.

\section*{Code availability}
All the codes that support the findings of this study are available from the corresponding author upon reasonable request.

\section*{Acknowledgments}
The work was supported by the German Research Foundation (DFG) via Kl-1439/14-1. M.F.K. is grateful for partial support by the Max Planck Society via the Max Planck Fellow program. A.F. and M.F.K.'s work at SLAC was supported by the U.S. Department of Energy, Office of Science, Basic Energy Sciences, under Contract No. DE-SC0063. W.Z. and J. W. acknowledge the funding support from the National Natural Science Foundation of China (Grant No.12304377, No. 12227807), and the Shanghai Pujiang Program (23PJ1402600). W.Z. and C.C.V. acknowledge support from the Alexander von Humboldt Foundation. A.S.C., S.A.M. and E.C. acknowledge funding and support from the Deutsche Forschungsgemeinschaft (DFG, German Research Foundation) under Germany’s Excellence Strategy -EXC2089/1-390776260, the Bavarian program Solar Energies Go Hybrid (SolTech), the Center for NanoScience (CeNS) and the European Commission through the ERC Starting Grant CATALIGHT (802989). A.S.C. acknowledges Xunta de Galicia, Spain, for her postdoctoral fellowship. We thank R.N. Shah (University of Freiburg) for helpful support during installation of the customized chirp mirror modules for the pump-probe setup.

\section*{Author Contributions} 
The experiments were set up by R.D. and W.Z., who carried out the experiments with support from P.R. R.D. and W.Z. analyzed the experimental data. The d-CTMC simulations were developed by P.R. and assessed and carried out by R.D.; T.M.L, A.N., P.V., F.S., designed the NAQMD simulations supported by discussions with R.D. and A.M.S.; T.M.L performed and analyzed the NAQMD simulations. A.S.C., S.A.M. and E.C. prepared and characterized the SiO$_2$ NP samples. M.N. provided support for laser operations. S.M. and S.B. designed and characterized the customized chirped mirror modules used for compressing the laser pulses. R.D., W.Z., P.R., C.C.V., B.B., and M.F.K. interpreted the data. R.D. and W.Z. drafted the manuscript with help of A.F. All authors participated in the discussion of the results and commented on the manuscript. M.F.K. and B.B. supervised the project and nanoTRIMS team.

\section*{Additional information}
Correspondence and requests for materials should be addressed to R.D., W.Z., or M.F.K.

\clearpage
\begin{center}
\Large Supplementary Materials
\end{center}
\subsection*{1. Materials and Methods}
\captionsetup[figure]{labelformat=empty}
\textbf{Experimental setup of pump-probe reaction nanoscopy:} 
\begin{figure}[h!]
\centering
\includegraphics[width=0.95\textwidth]{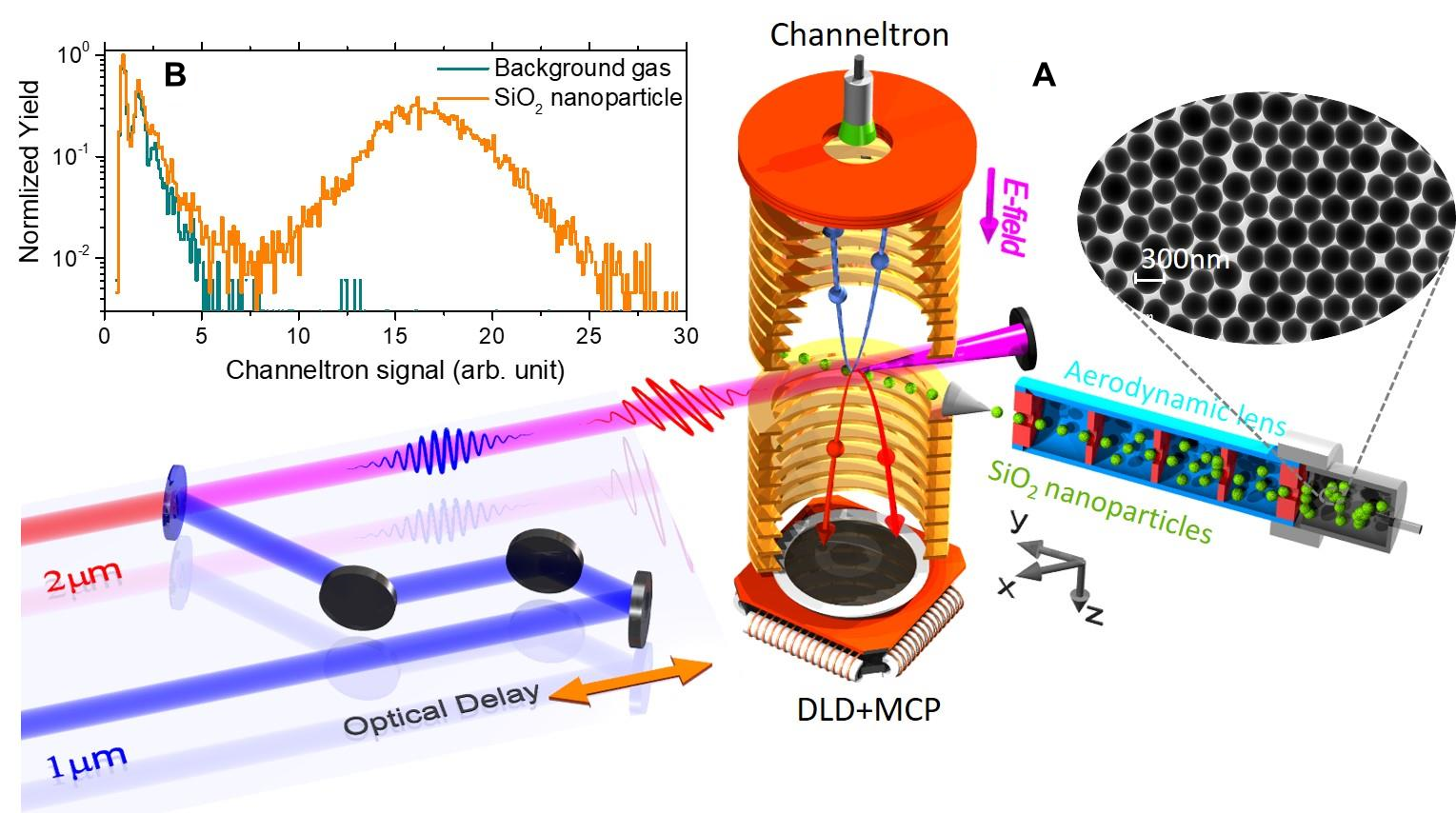}
\caption{\textbf{Fig. S1. Experimental scheme} (\textbf{A}) Schematic illustration of the femtosecond pump-probe experimental setup. The bottom-right inset shows a transmission electron micrograph (TEM) of 300 nm SiO$_2$ NPs used in the experiments. (\textbf{B}) Typical proton-correlated channeltron signals obtained from measurements of a target consisting of ethanol solvent with (orange curve) and without (dark-cyan curve) the presence of NPs.}
\end{figure}
As schematically illustrated in Fig. S1A, the experimental approach was the realization of time-resolved pump-probe measurements in the reaction nanoscope, also referred to as NanoTRIMS (Nano-Target Recoil Ion Momentum Spectroscopy) \cite{Rupp2019nc,Rosenberger2020,Zhang2022OPTICA,Rosenberger2022EJPD}. In the experiment, a mid-infrared pulse from an optical parametric chirped-pulse amplification laser system (17 fs, 2 \textmu m, 100 kHz) \cite{Neuhaus2018} was frequency doubled in a 1.5-mm-thick lithium niobate crystal. The two colors were separated in two arms of an interferometer using a dielectric mirror to produce a 2 \textmu m pump pulse and a 1 \textmu m probe pulse. Two customized chirped mirrors were installed in each arm to optimize the laser pulse duration (2 \textmu m, $\sim$ 35 fs; 1 \textmu m, $\sim$ 30 fs). The time delay between the pump and probe pulses was finely controlled by using a motorized delay stage (Physics Instruments, PI-M406.4PD). The linearly polarized pump and probe pulses were later recombined with a dielectric mirror. The collimated laser beams were tightly focused by a concave silver mirror (\textit{f} = 75 mm) inside the NanoTRIMS vacuum chamber, where they crossed the SiO$_2$ NP beam. The SiO$_2$ NPs of uniform size (304$\pm$19 nm) were synthesized through the process described in the following section. To produce the NP jet, NPs dispersed in pure ethanol with concentrations of 0.1 g/l, 0.2 g/l, and 0.5 g/l for 200 nm, 300 nm, and 600 nm particle sizes were aerosolized using a compressed gas atomizer (TSI, model 3076). The detailed method for NP synthesis and sample preparation can be found in a previous publication \cite{Zhang2022OPTICA}. The argon-carried NPs were dragged through a nafion-tubed membrane dryer (Perma Pure, PPMD-700-48S-1) to achieve a large removal of solvent. Afterward, the NP aerosol was introduced into the ultrahigh vacuum ($<$ 2×10$^{-10}$ mbar) chamber via an aerodynamics lens system, resulting in a collimated beam of isolated NPs streaming through the vacuum. 

Upon the interaction of the NPs with the focused strong laser pulses, the laser-created energetic ions and electrons ejecting from the NPs were guided by a homogenous electric field ($\sim$ 198 V/cm) to be detected in coincidence by two different detectors. A time- and position-sensitive detector consisting of a microchannel plate (MCP) stack and delay-line detector (DLD) (DLD80, RoentDek Handels GmbH) was utilized to measure the ion’s time-of-flight and impact positions from which the reconstruction of their full 3D momentum distributions was made. The hit rates of electrons associated with each laser shot were recorded using a channeltron electron amplifier (Photonis Magnum) at the opposite side of the NanoTRIMS spectrometer. As shown in Fig. S1B, a much higher channeltron signal was measured from NP-associated ionization events than that coming from background gas. The electrons and ion emission originating from molecular adsorbate ionization and dissociation events can thus be clearly distinguished. 

\textbf{Nanoparticle synthesis:} Tetraethylorthosilicate 98\% (TEOS), ammonium hydroxide solution 28-30\% (NH$_4$OH), pure grade ethanol were purchased from Sigma-Aldrich. All chemicals were used as received and Milli-grade water was used in all preparations. Monodisperse silica spheres (304$\pm$19 nm) were prepared using a modified St{\"o}ber method \cite{Stober1968}. 25 ml of EtOH solution containing 9.17 ml of H$_2$O, 2.44 ml NH$_4$OH and 1.42 ml of TEOS was quickly added and the mixture was kept for 2 hours. The resultant SiO$_2$ nanospheres were washed with ethanol by three centrifugation-redispersion cycles (5000rpm, 15 min).

\subsection*{2. Simulation details}
\textbf{d-CTMC Simulations.} The basis of the d-CTMC simulations is similar to the static field model as used in references \cite{Rupp2019nc, Rosenberger2020, Zhang2022OPTICA, Rosenberger2022EJPD} and is adapted to capture the time-dependent surface charge dynamics. To elaborate, the near-field distribution around silica spheres with a given and refractive index of 1.5, was calculated using the Mie solution numerically. The initial local surface charges are generated using the Monte-Carlo rejection sampling technique based on the ADK \cite{Ammosov1986} ionization rate as point charges on the particle surface according to the near-field-dependent ionization probability distribution for a 2\textmu m laser pulse. For the ADK calculations, we used the maximum peak electric field above the NP surface.  The protons’ birth positions for the first laser pulse were sampled according to the dissociation ionization probabilities  modeled by a power law $I^{n}$, specifically using $n = 26$, ensuring consistency with the 2 \textmu m laser wavelength employed in this study, multiplied by a radial Gaussian distribution ($r_0$ = 150 nm, $\sigma_{r}$ = 30 nm ) to account for the spatial distribution of the molecules. We note that once the pump pulse has subsided, it leaves a highly positively charged NP surface. The post-laser "equilibrium-phase" of the surface charge redistribution is considered to be dominated by the diffusion of charges and charge loss. To model that, we diffused the mother ions sampled on the surface by a Gaussian random walk along the surface, also reckoning the charge loss mechanism via the exponential decay equation:
\begin{equation}
    Q(t) = Q_0 e^{-kt}+Q_{residual}
\end{equation}
    
Here, $Q(t)$ denoted the number of surface charges remaining at time $t$, $Q_0$ is the initial amount of charge, $k$ is the charge decay constant that characterizes the number of charges decreasing over time, and $Q_{residual}$ is the residual charge which represents the remaining charge on the surface after the time window for the simulation has lapsed. The 3D random walk is governed by a diffusion coefficient, where the step size is drawn randomly from a Gaussian distribution multiplied by $\sqrt{6D\tau}$ expressing the mean displacement with a time step $\tau$, moving in random directions. Here, $D$ denotes the diffusion coefficient. The total number of steps is calculated as by dividing the time interval by $\tau$ = 1 ps (also tested for smaller steps of 500 fs leading to the same results), for each progressing time step. During each diffusion step, we enabled the sampled protons to undergo classical propagation away from the surface, influenced by the electrostatic field of the surface charges according to Newton's law to obtain the final 3D momentum spectra neglecting the effect of the fast-escaped electrons bunch on proton dynamics. We then gathered the resulting 3D momentum spectra of the protons at each step \cite{Rosenberger2020, Zhang2022OPTICA}. The diffusion coefficient, 1.35 cm$^2$/s, and the charge decay constant approximated as 40 ns, were adjusted as fitting parameters to align the decay behavior with the experimental results. In the context of proton sampling, we postulated that the probability of proton generation is influenced by the presence of a surface charge, which facilitates bond weakening. To address the non-linear distribution of the local charge density, and thus, proton sampling positions, we adopted an initial proton sampling region with a power-law dependence. This initial region was subsequently subjected to diffusion, reflecting the dynamics of surface charges. For the simulations, we introduced a second pulse to trigger the dissociation of weakened bonds in the presence of the high electric field-induced polarization of the NP. As a result, proton generation was found to be enhanced in regions with higher local surface charge density. 
	
 In the model, it is assumed that the protons emitted with an initial velocity of zero are propagated independently of each other and the emission dynamics is dominated by the electrostatic interactions with the positively charged NP surface. The effect of Coulomb interaction from the fast escaping electrons on the final proton momenta is assumed to be negligible. In the model, the number of sampled static surface charges for the normalization of initial charge distribution $\rho_\mathrm{charge}$ was estimated to be $\sim$ 1810\textit{e} by fitting the single-2\textmu m-induced proton KE distribution at an intensity of 4 $\times$ 10$^{13}$ W/cm$^2$.

\vspace{18pt}
\textbf{NAQMD Simulations.} To examine the role of charge generation on bond softening and proton dissociation in the silica nanosphere, we adopted a multiscale approach, as first principle simulations are unable to simulate surfaces on the size of the NP. We apply a strong DC field along a silica surface to model the effect of the fields induced by the global hole density on the nanosphere similar to globally informed Hartree-potential frameworks in divide and conquer density functional theory.  Thus, the simulated silica surface represents a $\sim$1nm$^2$ chunk of silica embedded in an ionized nanosphere.  To obtain an order of magnitude estimate of the fields induced on a nano-sphere from a non-uniform ionized charge density, we first utilize a Mie field solver to obtain the near field enhancement at the nano-sphere surface for a given pump pulse. From the local optical fields, we generate an ionized charge distribution utilizing Ammosov, Delone, and Krainov (ADK) theory \cite{Ammosov1986}. The integrated ADK rate is used to determine the probability density of ionization for any given surface area element, and the total ionized charge density is the set as density of “SiO$_2$” in crystalline silica times the ionization probability \cite{Seiffert2018}. Fig. S2A shows the charge distribution at the surface as function of the polar angle $\theta$ for an intensity 5 $\times$ 10$^{13}$ W/cm$^2$. To obtain the electric field induced by the charge density we solve Poisson’s equation via a multipole expansion and the field at the NP surface is plotted in Fig. S2B. While many approximations were made to obtain this field, the goal was to inform boundary conditions for Non-Adiabatic Quantum Molecular (NAQMD) dynamics simulations for which only an order of magnitude estimate of expected electric fields generated in the experiments is required. 

\begin{figure}[t]
\centering
\includegraphics[scale=0.12]{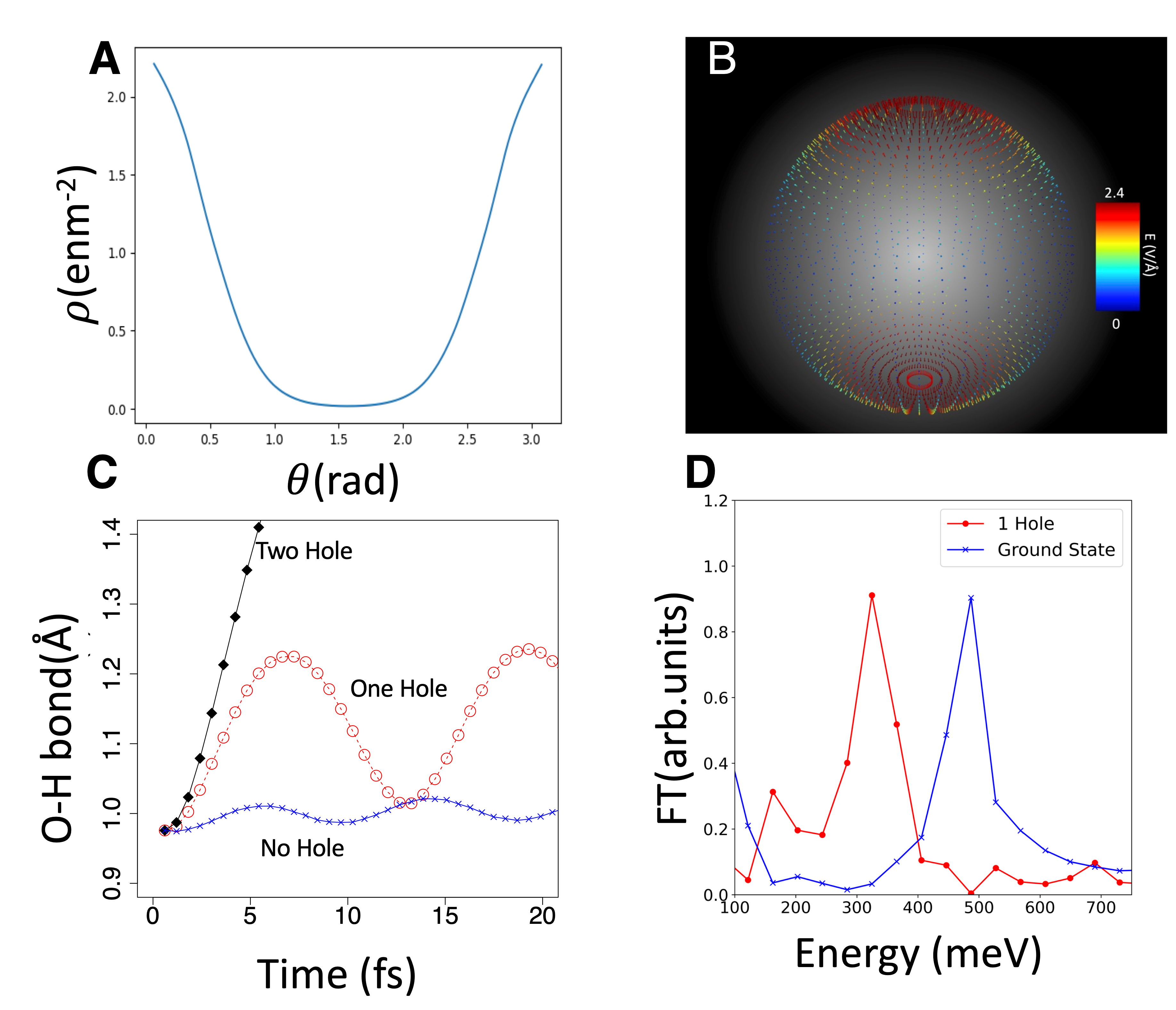}
\caption*{\textbf{Fig. S2. NAQMD Simulations.} (\textbf{A}) charge distribution on the surface as a function of polar angle $\theta$ at the nanosphere (\textbf{B}) Calculated electric field at the surface of the NP generated by the charge density obtained via using Poison's equation through a multipole expansion (\textbf{C}) Dynamics of an O-H bond on the surface, focusing on the hole wave-functions with the highest Mulliken population (\textbf{D}) Decrease in vibrational energy associated with the O-H stretching mode, as observed in the Fourier transform of the O-H bond length dynamics.}
\end{figure}

NAQMD simulations were performed on a 14.32Å cubic length amorphous silica structure. Along the one of the cubic axes, we created a hydrogen terminated surface and added additional vacuum to prevent image interactions. An electric field of 0.02 atomic units ($\sim$1V/Å) was applied using a sawtooth potential. NAQMD simulations are an ab-intio molecular dynamics approach that integrates the trajectories of all atoms by computing their intermolecular forces from first-principles in the framework of density functional theory(DFT).To approximate the exchange-correlation functional in our DFT approach we used the Perdew-Burke-Ernzerhof (PBE) version of generalized gradient approximation (GGA) \cite{Perdew1996}. Van der Walls corrections were employed utilizing the DFT-D scheme \cite{Grimme2010}. The projected argument wavevector (PAW) method was used to calculate electronic states within a plane wave basis set \cite{Blochl1994}. Projector functions were generated for the Si 3s and 3p states, O 2s and 2p states, and the 1s state for hydrogen. A planewave cutoff energy of 35 Ry was used. NAQMD allows for dynamics of excited carriers to be modeled within the framework of time dependent DFT. Excited state transitions were modeled within the fewest switch surface hopping method \cite{Tully1990}. Dynamics were performed in the microcanonical ensemble at 300K. The NAQMD algorithm was implemented in QXMD quantum molecular dynamics simulation code \cite{Shimojo2019}. For further details of the NAQMD algorithm we refer the readers to references \cite{Shimojo2019, Fuyuki2013, Fuyuki2014}. 

Fig. S2C shows dynamics of an O-H bond on the surface for which the hole wave-functions had the highest Mulliken population. For two holes excited from the valence band maximum (VBM) the O-H bond quickly dissociates. With one hole excited a clear bond-softening is seen in comparison to the field and hole free ground state. The decline in O-H stretching vibrational energy is illustrated in Fig. S2D taken from Fourier transform of the O-H bond length dynamics. This softening can lead to breaking during the stretching of the bond by the probe pulse.

\subsection*{3. Measured proton momenta at negative delays}
\begin{figure}
\centering
\includegraphics[scale=0.9]{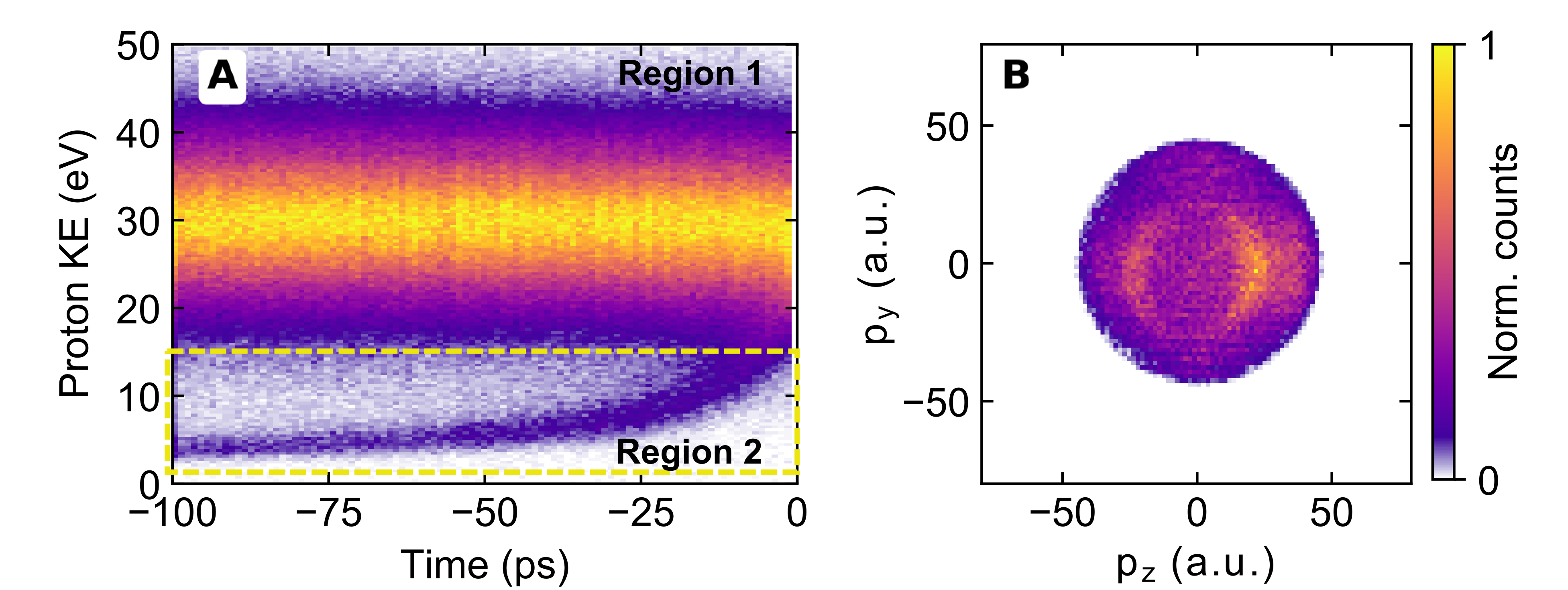}
\caption*{\textbf{Fig. S3. Distribution of protons measured for negative delays}, similar to Figure 2 in the main text.  (\textbf{A}) Measured proton KE spectra as a function of pump-probe time delay. Here, the 1 \textmu m laser pulse assumes the role of the pump, while the 2 \textmu m laser pulse functions as the probe.  (\textbf{B}) Time-integrated proton momentum distribution associated with the KE decay signal in region 2 of the panel (A).}
\end{figure}

As mentioned in the main text, the behavior of protons in response to changes in the role of the pump and probe mirrors the patterns depicted in Fig. S3 on the negative delay side. In this context, the 1 \textmu m laser pulse assumes the role of the pump, while the 2 \textmu m laser pulse functions as the probe. Note that the delay range extends from 0 to 100 ps for negative delays. Fig. S3A presents the observed distribution of proton kinetic energy (KE) as a function of delay, with the roles of pump and probe being reversed between the 2 \textmu m and 1 \textmu m pulses in comparison to the data described in the primary text. In this particular setup, we observe that the emission directions of protons now align with the polarization direction of the 1 \textmu m laser pulse i.e., along the z-direction, which serves as the pump in these results. This contrasts with the observations in Fig. 2E, where the dipolar emission of protons, associated with the KE decay signal, followed the polarization direction of the 2 \textmu m pulse, which was in the y-direction. It's worth noting that the weaker intensity of the 1 \textmu m pulse only leads to a decrease in statistics for the overall time-dependent proton signal, while the underlying mechanism related to proton dissociation through bond-weakening remains unaltered. This consistency lends support to the interpretation that the bond-weakening process on the surface is primarily driven by charge-mediated mechanisms.

\subsection*{4. Nanoparticle-size effect on charge dynamics}
Here, the impact of particle size on charge dynamics is explored. Proton yields are measured for NPs of different sizes (200 nm, 300 nm, and 600 nm) with varying kinetic energy (KE) spectra and pump-probe time delays as demonstrated in Fig. S4A-C. Although proton KE spectra and their time-dependence appear similar across different particle sizes, closer examination reveals a variation of time constants in relation to the bi-exponential fit represented by the equation $y = y_0 + A_1[\exp(-t/\tau_{fast})] + A_2[\exp(-t/\tau_{slow})]$, where $A_{1,2}$ denote the amplitude, $y_0$ stands for the offset, and $\tau_{fast}$ and $\tau_{slow}$ represent the time constants. The obtained values of the time constants are $\tau_\mathrm{fast} = 8.8$ $\pm$ 0.2 ps, $\tau_\mathrm{slow}$ = 65.02 $\pm$ 3.2 ps for 200 nm, $\tau_\mathrm{fast} = 10.3$ $\pm$ 0.1 ps, $\tau_\mathrm{slow} = 70.1$ $\pm$ 1.8 ps for 300 nm, and $\tau_\mathrm{fast} = 15.0$ $\pm$ 0.4 ps, $\tau_\mathrm{slow} = 80.02$ $\pm$ 2.5 ps for 600 nm particles, respectively. \begin{figure}
\centering
\includegraphics[width=1\textwidth]{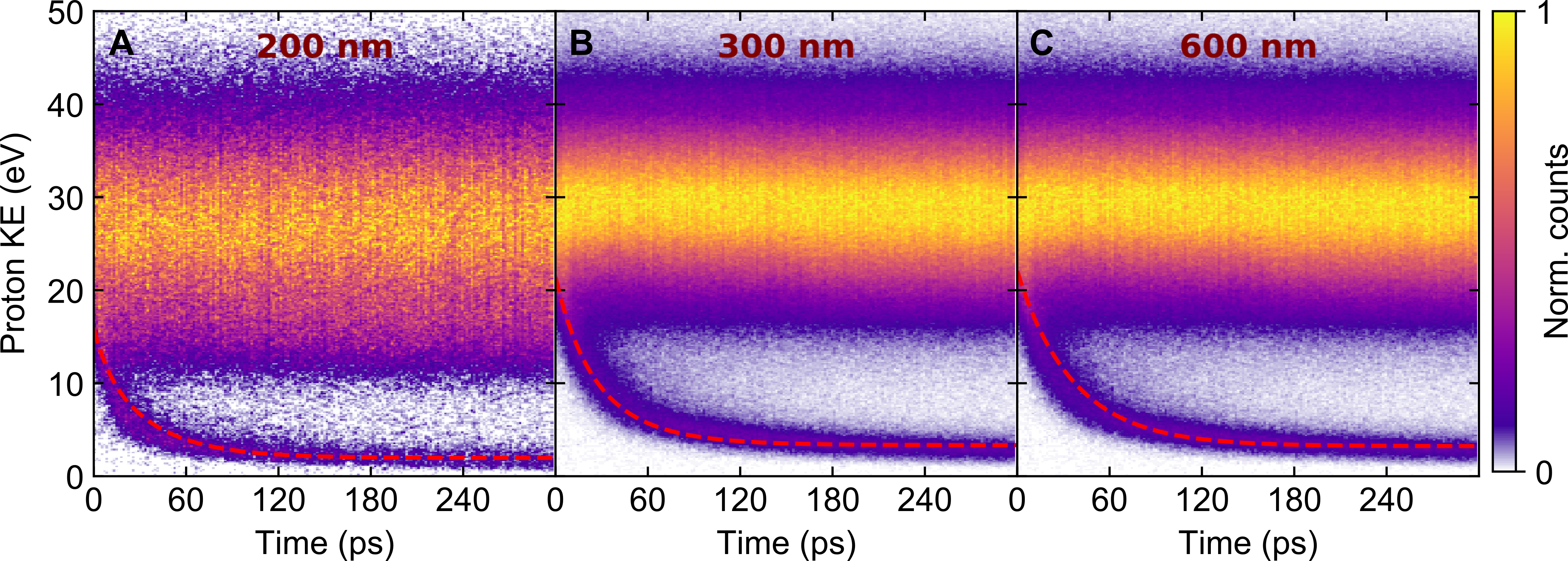} 
\caption{ \textbf{Fig. S4. NP size effects on the proton KE decay.} Measured proton KE distributions as a function of time delay for (\textbf{A}) 200 nm, (\textbf{B}) 300 nm, and (\textbf{C}) 600 nm, particles, respectively. The red-dashed curves on top of the experimental data correspond to theoretically calculated mean KE decay values for each particle size.}
\end{figure}

The timescales for the fast and slow processes of surface charges are shorter for smaller NPs and become larger as size increases showing that smaller NPs exhibited faster charge redistribution dynamics in contrast to their larger counterparts. This can be attributed to their reduced surface areas, highlighting the significant influence of particle size on these dynamics. At any given time, the dispersion of laser-induced surface charges on smaller NPs yields a broader surface distribution characterized by lower charge density, thus contributing to a lower final proton KE. To confirm the reliability of the theoretical constants obtained for a specific size, i.e., \textit{d} = 300 nm SiO$_2$, the d-CTMC simulations extended to include NPs of other sizes (\textit{d} = 200 nm, 600 nm), resulted in remarkably consistent outcomes. These results are depicted in Fig. S4A-C, superimposed over the experimentally measured data represented by red-dashed curves. This indicates that the surface charge redistribution occurs quickly for smaller NPs owing to their smaller surface areas while the timescales increase as the NP surface area increases taking longer to achieve the same local surface charge densities.
\begin{figure}
\centering
\includegraphics[width = 1\textwidth]{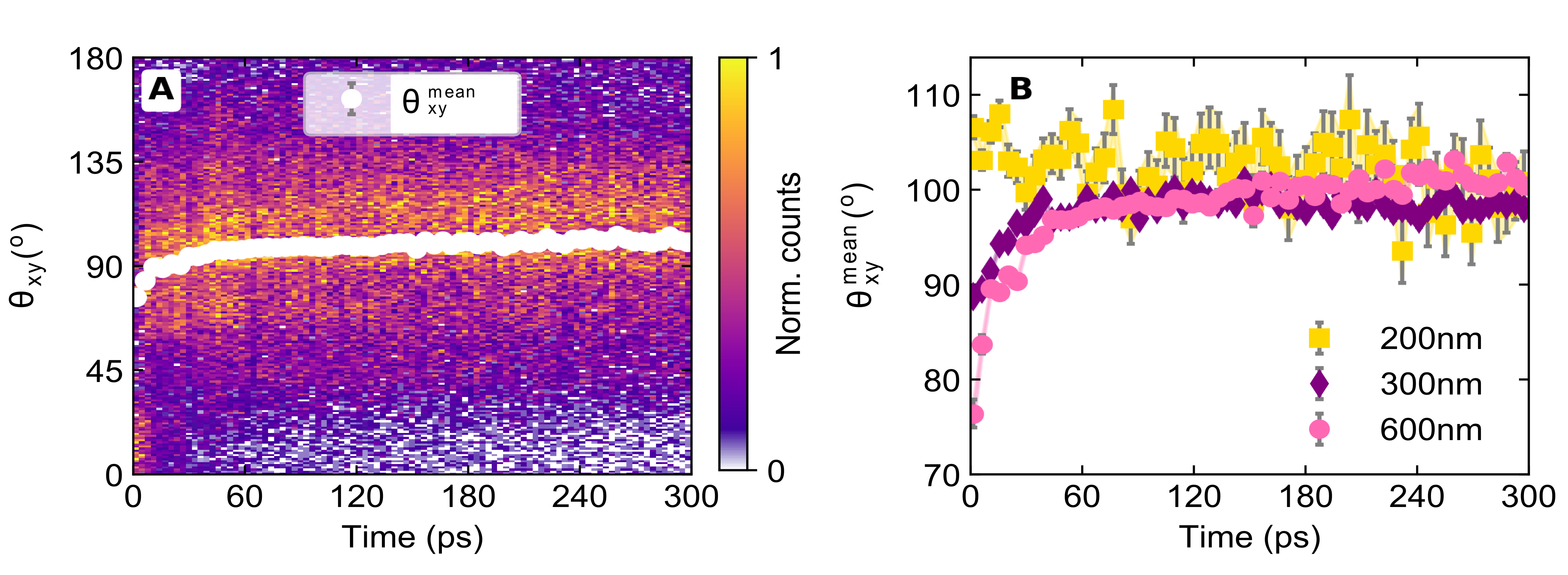}
\caption*{\textbf{Fig. S5. Field-propagation effects.} (\textbf{A}) The proton angular distribution $\theta_{xy}$ (within the laser-polarization-propagation plane) is presented as a function of time delay for 600 nm particles. The white solid circles represent the mean values of the angular distribution in $\theta$ at different times. (\textbf{B}) Comparison of the mean values of the angular distribution in $\theta$ as a function of time delay presented for 200 nm, 300 nm, and 600 nm particles, as indicated.}
\end{figure}

Also, larger NPs experience significant deflection in proton emission angles due to the field-propagation effect \cite{Süßmann2015nc}, resulting in a bias against forward laser direction. A size-dependent field-propagation effect on surface charge dynamics is also observed as shown in the 2D density map of the change in the mean angle of the distribution of $\theta$ in the propagation-polarization (x-y) plane as a function of time in Fig. S5A for 600 nm SiO$_2$ spherical NPs. A size-dependent comparison of the mean of the angular distribution $\theta$ is shown in Fig. S5B for different sizes of NPs as time increases. The effect is more pronounced for larger particles, leading to more substantial deflection. The deflection trends plateau when charges reach the poles. This interplay of particle size and field-propagation effects on charge dynamics sheds light on how the light-matter interaction determines the initial charge distribution, which in turn affects the subsequent dynamics.

\subsection*{5. H$_2^+$ emission}
In addition to the processes detailed in the main text, which primarily relied on observing the emission of protons as an indicator of the NP surface characteristics, we aim to delve more into the chemistry taking place at the NP surface here. The TOF measurements depicted the record of H$_2^+$,  H$_3^+$ fragments along with H$^+$ ions. In a previous study\cite{Alghabra2021}, reaction nanoscopy identified the role of SiO$_2$ NPs as catalysts in the formation of trihydrogen from water molecules residing on the NP surface through bimolecular photoreactions. In our current study, employing ethanol as the solvent for NP dispersion, the studies reveals that the H$_2^+$ and H$_3^+$ fragments originate from the -O-CH$_2$CH$_3$ (-ethoxy) groups attached to the surface. While the formation of H$_2^+$ and H$_3^+$ fragments from organic molecules has been previously studied\cite{Ekanayake2018}, in our case the charged NP surface enhances these reactions and yields highly energetic fragments in comparison.

Figure S6 illustrates the recorded time-dependent kinetic energy spectrum of H$_2^+$ ions. Notably, H$_2^+$ ions display a kinetic energy decay pattern similar to that of protons. However, they possess lower average kinetic energies than protons, which can be attributed to the energy expended during the formation of H$_2^+$ ions. The disparities in the masses of these fragments can also influence their interactions with the charged NP surface. The fast component of the bi-exponential fit applied to the kinetic energy decay curve yields a time constant of 25.3 ps. This suggests that H$_2^+$ ions escape from the NP surface at a slower rate. This deceleration can be linked to the heavier mass of H$_2^+$ ions, causing protons to escape more rapidly in comparison. 

\begin{figure}
\centering
\includegraphics[width=\textwidth]{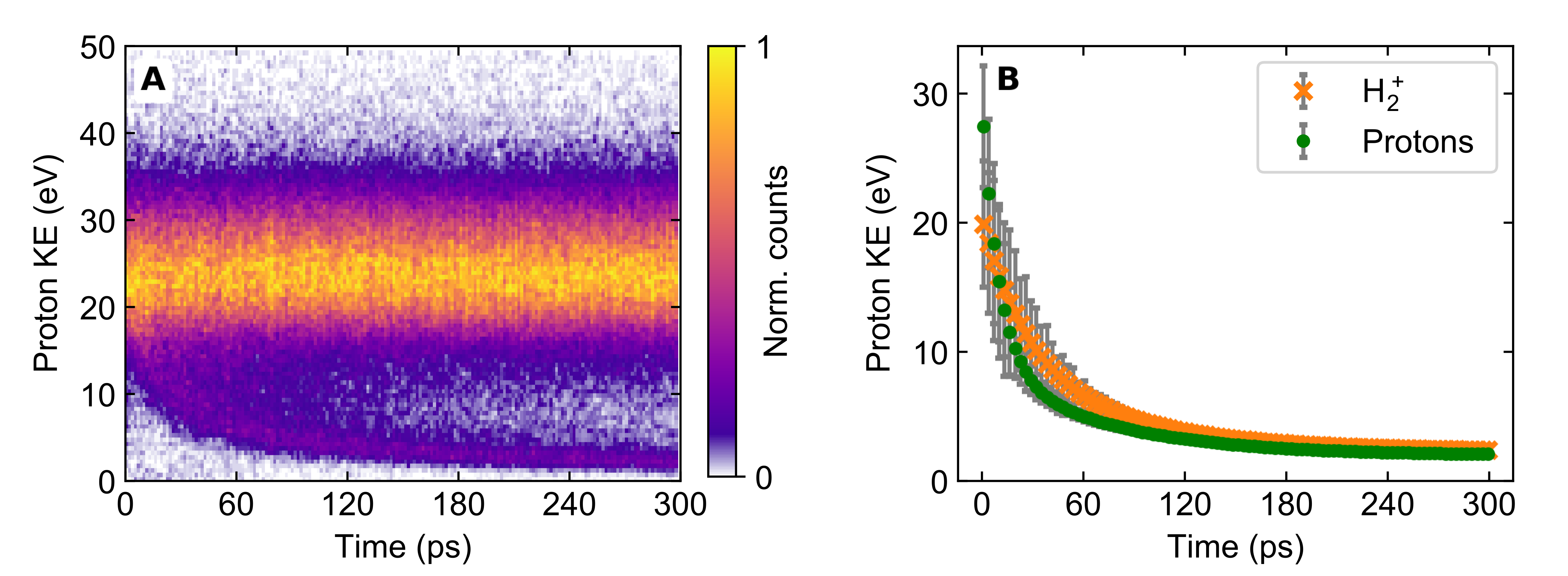}
\caption*{\textbf{Fig. S6. H$_2^+$  fragment dynamics.} (\textbf{A}) H$_2^+$ KE distributions as a function of the time delay measured for 300 nm. The time delay was scanned from 0 to 300 ps. (\textbf{B}) Time-sliced mean values of the delay-dependent  KE decay signal for H$_2^+$ and protons. The solid curves are the bi-exponential fits to the data points.}
\end{figure}

\end{document}